\documentclass{aastex631}

\shorttitle{Prediction of Halo Coronal Mass Ejections}
\shortauthors{Zhang et al.}

\begin{document}

\title{Prediction of Halo Coronal Mass Ejections Using 
SDO/HMI Vector Magnetic Data Products and a Transformer Model}

\correspondingauthor{Jason Wang, Haimin Wang}
\email{wangj@njit.edu, haimin.wang@njit.edu}

\author{Hongyang Zhang}
\affiliation{Institute for Space Weather Sciences, New Jersey Institute of Technology, University Heights, Newark, NJ 07102, USA} 
\affiliation{Department of Computer Science, New Jersey Institute of Technology, University Heights, Newark, NJ 07102, USA}

\author{Ju Jing}
\affiliation{Institute for Space Weather Sciences, New Jersey Institute of Technology, University Heights, Newark, NJ 07102, USA}
\affiliation{Center for Solar-Terrestrial Research, New Jersey Institute of Technology, University Heights, Newark, NJ 07102, USA}
\affiliation{Big Bear Solar Observatory, New Jersey Institute of Technology, 40386 North Shore Lane, Big Bear City, CA 92314, USA}

\author{Jason T. L. Wang}
\affiliation{Institute for Space Weather Sciences, New Jersey Institute of Technology, University Heights, Newark, NJ 07102, USA}
\affiliation{Department of Computer Science, New Jersey Institute of Technology, University Heights, Newark, NJ 07102, USA}

\author{Haimin Wang}
\affiliation{Institute for Space Weather Sciences, New Jersey Institute of Technology, University Heights, Newark, NJ 07102, USA}
\affiliation{Center for Solar-Terrestrial Research, New Jersey Institute of Technology, University Heights, Newark, NJ 07102, USA}
\affiliation{Big Bear Solar Observatory, New Jersey Institute of Technology, 40386 North Shore Lane, Big Bear City, CA 92314, USA}

\author{Yasser Abduallah}
\affiliation{Institute for Space Weather Sciences, New Jersey Institute of Technology, University Heights, Newark, NJ 07102, USA}
\affiliation{Department of Computer Science, New Jersey Institute of Technology, University Heights, Newark, NJ 07102, USA}

\author{Yan Xu}
\affiliation{Institute for Space Weather Sciences, New Jersey Institute of Technology, University Heights, Newark, NJ 07102, USA}
\affiliation{Center for Solar-Terrestrial Research, New Jersey Institute of Technology, University Heights, Newark, NJ 07102, USA}
\affiliation{Big Bear Solar Observatory, New Jersey Institute of Technology, 40386 North Shore Lane, Big Bear City, CA 92314, USA}

\author{Khalid A. Alobaid} 
\affiliation{Institute for Space Weather Sciences, New Jersey Institute of Technology, University Heights, Newark, NJ 07102, USA}

\affiliation{College of Applied Computer Sciences, King Saud University, Riyadh 11451, Saudi Arabia}

\author{Hameedullah Farooki}
\affiliation{Department of Astrophysical Sciences,
Princeton University, Princeton, NJ 08544, USA} 

\author{Vasyl Yurchyshyn}
\affiliation{Big Bear Solar Observatory, New Jersey Institute of Technology, 40386 North Shore Lane, Big Bear City, CA 92314, USA}

\begin{abstract}
We present a transformer model, named DeepHalo, to predict
the occurrence of halo coronal mass ejections (CMEs).
Our model takes as input an active region (AR) 
and a profile,
where the profile contains a time series of data samples in the AR that are collected 24 hours
before the beginning of a day,
and predicts whether the AR would produce a halo CME
during that day.
Each data sample contains 
physical parameters, 
or features,
derived from photospheric vector magnetic field data taken by the Helioseismic and Magnetic Imager (HMI) on board the
Solar Dynamics Observatory (SDO).
We survey and match CME events
in the Space Weather Database Of Notification, Knowledge, Information (DONKI)
and Large Angle and Spectrometric Coronagraph
(LASCO) 
CME Catalog, and compile a list of CMEs
including halo CMEs and non-halo CMEs
associated with ARs in the period 
between November 2010 and August 2023.
We use the information gathered above to build the labels (positive versus negative) of the
data samples and profiles at hand, where the labels are needed for machine learning.
Experimental results
show that DeepHalo with a
true skill statistics (TSS) score of 
0.907 outperforms
a closely related long short-term memory network
with a TSS score of
0.821.  
To our knowledge, this is the first time that the transformer model has been used for halo CME prediction.
\end{abstract}

\section{Introduction} 
\label{sec:intro}

Coronal mass ejections (CMEs)
propel the mass and magnetic substances
from the Sun into interplanetary space
on a rapid timescale
\citep{1982ApJ...263L.101H,
2000JGR...105.2375L,
Gopal2005JGRA,
2012hssr.book.....S,
WH2012LRSP,
Pal2018ApJ,
2019shin.confE..92G,
KLM2019SpWea,
2020SpWea..1802478U,
MDT2022A&A}. 
CMEs toward Earth have the potential to disrupt crucial technologies on Earth or in the near-Earth environment, 
including satellite systems, communication systems, power grid systems, 
and many more
\citep{2004SpWea...2.2004B,2012hssr.book.....S}.
Given the potential threats posed by CMEs, 
significant efforts have been made to improve technologies
for the early detection and forecasting of CMEs and to
estimate their magnetic fields, characteristics, properties, and travel time
\citep[e.g.,][]{2008SoPh..248..471Q,
BI2016ApJ,
2016EGUGA..18.4784P,
2018ApJ...855..109L, 
2019SoPh..294..130K,
2019ApJ...881...15W,  
2020ApJ...890...12L,
2022cosp...44.1405K,
2023ApJ...958L..34A,
Guastavino2023}. 
Among the CMEs, 
halo CMEs including full-halo CMEs and partial-halo CMEs
are of particular importance
due to their Earth-directed propagation tendencies
\citep{2006ApJ...646.1335L,2009EP&S...61..595G}. 
Full-halo CMEs exhibit an apparent complete angular width
of precisely $360^{\circ}$. 
Partial-halo CMEs manifest with angular width $(W)$ where
$120^{\circ} \leq W < 360^{\circ}$
\citep{2006SpWea...410003M,2007JGRA..112.6112G,2009EP&S...61..595G}.
Such CMEs are commonly associated with
solar flares and erupting filaments
\citep{2009EP&S...61..595G}.

Instruments such as the Large Angle and Spectrometric Coronagraph (LASCO) 
on board the Solar and Heliospheric Observatory \citep[SOHO;][]{1995SoPh..162....1D} 
and the Helioseismic and Magnetic Imager (HMI)
on board the Solar Dynamics Observatory 
\citep[SDO;][]{2012SoPh..275..207S} 
have been devised to make observations of the Sun and near-Sun environment.
Researchers have used SDO/HMI data products, such as Space-Weather HMI Active Region Patches \citep[SHARPs;][]{2014SoPh..289.3549B}
and other active region (AR) properties, to forecast solar eruptions, including CMEs
\citep[e.g.,][]{2015ApJ...798..135B,BI2016ApJ,
2017ApJ...843..104L,
2018ApJ...861..128I,
2019ApJ...883..150C,
2019SpWea..17.1404C,
2019ApJ...885...73Z,
2020ApJ...898...98B,
2020ApJ...891...10L,
2020ApJ...890...12L,
2020ApJ...890..124P,
2020ApJ...895....3W,
2021ApJS..257...50T,
2022ApJ...941....1S,
2022ApJS..263...28Z,
2023NatSR..1313665A}.
The AR patches provide detailed magnetic- and velocity-field information, giving us a closer look at the Sun's activities. 
The valuable insights they offer shed light on the dynamics and
intricacies of solar ARs,  
from which most CMEs originate \citep{2019RSPTA.37780094G},
although some CMEs originate from quieter regions
\citep{1996JGR...10113497M,2000JASTP..62.1427S}.

In this paper, we attempt to
use SHARP magnetic parameters to predict
the occurrence of a special class of CMEs, namely halo CMEs, emitted from ARs.
Previous studies on CME forecasting
confined their investigations to flaring ARs
\citep{BI2016ApJ,2020ApJ...890...12L}.
However, solar observational records over the years
suggest that there may not be a direct one-to-one correspondence between flares and CMEs \citep{2009IAUS..257..283G}, 
although they often occur together. 
Thus, unlike the previous methods,
we develop a transformer model, named DeepHalo, to predict whether an AR would produce
a halo CME regardless of whether the AR is flaring or not.
Our work will help scientists better understand and forecast the
geoeffectiveness of CMEs, 
as halo CMEs have a high potential to cause geomagnetic storms
\citep{2006SpWea...410003M,2007JGRA..112.6112G,2009EP&S...61..595G}.

The transformer is a deep learning model
that stands out with its inherent attention mechanism, allowing it to meticulously weigh different segments of input data \citep{DBLP:conf/nips/VaswaniSPUJGKP17}. 
It is particularly suitable for sequential data sets with applications ranging from speech analysis
to time-series prediction.  
Compared to other deep learning models such as long short-term memory (LSTM) networks \citep{2020ApJ...890...12L},
the transformer exhibits marked superiority in capturing mid- to long-range dependencies
within time series data. 
Unlike LSTM networks, which process time series in a sequential order and may struggle to capture long-distance dependencies, the transformer leverages its self-attention mechanism to
directly capture the dependencies between any two points in a sequence. 
This capability eliminates the need to accumulate information in multiple time steps,
thereby enhancing the transformer's efficiency and accuracy in time-series forecasting. 

The remainder of this paper is organized as follows. 
Section \ref{sec:data} describes our data collection scheme
and the predictive parameters used in this study. 
Section \ref{sec:method} defines the prediction task at hand
and presents our transformer model and algorithms to tackle the task.
Section \ref{sec:results} reports the experimental results.  
Section \ref{sec:conclusion} presents a discussion and concludes the paper.

\section{Data} \label{sec:data}

We considered CMEs in the Space Weather Database Of Notification, 
Knowledge, Information 
(DONKI)\footnote{\url{http://kauai.ccmc.gsfc.nasa.gov/DONKI/}}
and LASCO CME Catalog.\footnote{\url{https://cdaw.gsfc.nasa.gov/CME_list/}}
We surveyed CMEs that occurred in the period between November 2010 and
August 2023 and compiled a list of CME events 
associated with the ARs in the period. 
This period was chosen because of the availability of consistent and comprehensive data from the LASCO CME Catalog
and DONKI since the launch of SDO in 2010.
We matched the CME list between the LASCO CME Catalog and DONKI. 
This matching process ensured the accuracy of our data, as we cross-referenced CME events from two data sources. 
If a CME's start time from DONKI is within the two-hour window of a CME's first C2 appearance
time in the LASCO CME Catalog, 
we treat the CMEs as the same event.
This two-hour window is set to account for minor discrepancies in the recording times between the two data sources.
The data collection process yields a data set
with 1283 CMEs, in which
117 are full-halo CMEs,
209 are partial-halo CMEs, 
and 957 are non-halo CMEs. 
Non-halo CMEs manifest with angular width ($W$)
where $W < 120^{\circ}.$

In addition, we used the SHARP data products provided by the SDO/HMI team
as input to our transformer model.
These data products are publicly available from
the Joint Science Operations Center
(JSOC).\footnote{\url{http://jsoc.stanford.edu}} 
Diving into the SHARP data, 
one can find meticulously charted magnetic field maps of
every distinct solar AR since the SDO's advent. 
These ARs, captured in their entirety, are accompanied by 18 descriptive parameters that encapsulate the intricacies of photospheric magnetic fields \citep{2014SoPh..289.3549B}. 
Parameters related to
energy and magnetic flux include
the mean photospheric magnetic free energy (MEANPOT), the total photospheric magnetic free energy density (TOTPOT), the aggregate unsigned flux (USFLUX), 
and the accumulated flux near the polarity inversion line (R\_VALUE). 
Parameters related to
helicity and current dynamics
include the average current helicity (MEANJZH), 
the absolute magnitude of the net current helicity (ABSNJZH), 
the total modulus of net current across polarities (SAVNCPP), 
the entire unsigned vertical current (TOTUSJZ) 
and the complete unsigned current helicity (TOTUSJH). 
Additionally, the prevalent vertical current density (MEANJZD) provides information on the magnetic currents in a region. 
Another significant aspect revolves around the twist, shear, and angular orientation of magnetic fields. 
Parameters such as the average characteristic twist (MEANALP), 
the prevalent shear angle (MEANSHR), 
the proportion of the area with
a shear angle exceeding 45$^{\circ}$ (SHRGT45) 
and the average deviation of the magnetic field from the radial direction (MEANGAM) fall into this category. 
Lastly, the gradient and strength of magnetic fields play a pivotal role in understanding solar activity. 
They can be deciphered from parameters
such as the average gradient of vertical (MEANGBZ), 
total (MEANGBT), and horizontal (MEANGBH) magnetic fields. 
Additionally, the region's pixel area 
characterized by intense magnetic fields (AREA\_ACR) 
offers a perspective on the spatial extent of magnetic interactions. 
Each data sample used in our study contains all 18 magnetic parameters, or features,
at a cadence of 12 minutes.

Because the magnetic parameters have different units and scales, we
normalize the parameter values as follows.
Let $z_i^k$ denote the normalized value of the $i^{th}$ parameter of the $k^{th}$ data sample. Then
\begin{equation}
z_i^k=\frac{v_i^k-\mu_i}{\sigma_i},
\label{normalization-1}
\end{equation}
where $v_i^k$ is the original value of the $i^{th}$ parameter of the $k^{th}$ data sample, 
$\mu_i$ is the mean value of the $i^{th}$ parameter, and $\sigma_i$ is the standard deviation of the $i^{th}$ parameter. 

\section{Methodology} \label{sec:method}
\subsection{Prediction Task} 
\label{subsec:task}

\begin{figure}
\centering
\includegraphics[width=0.9\linewidth]{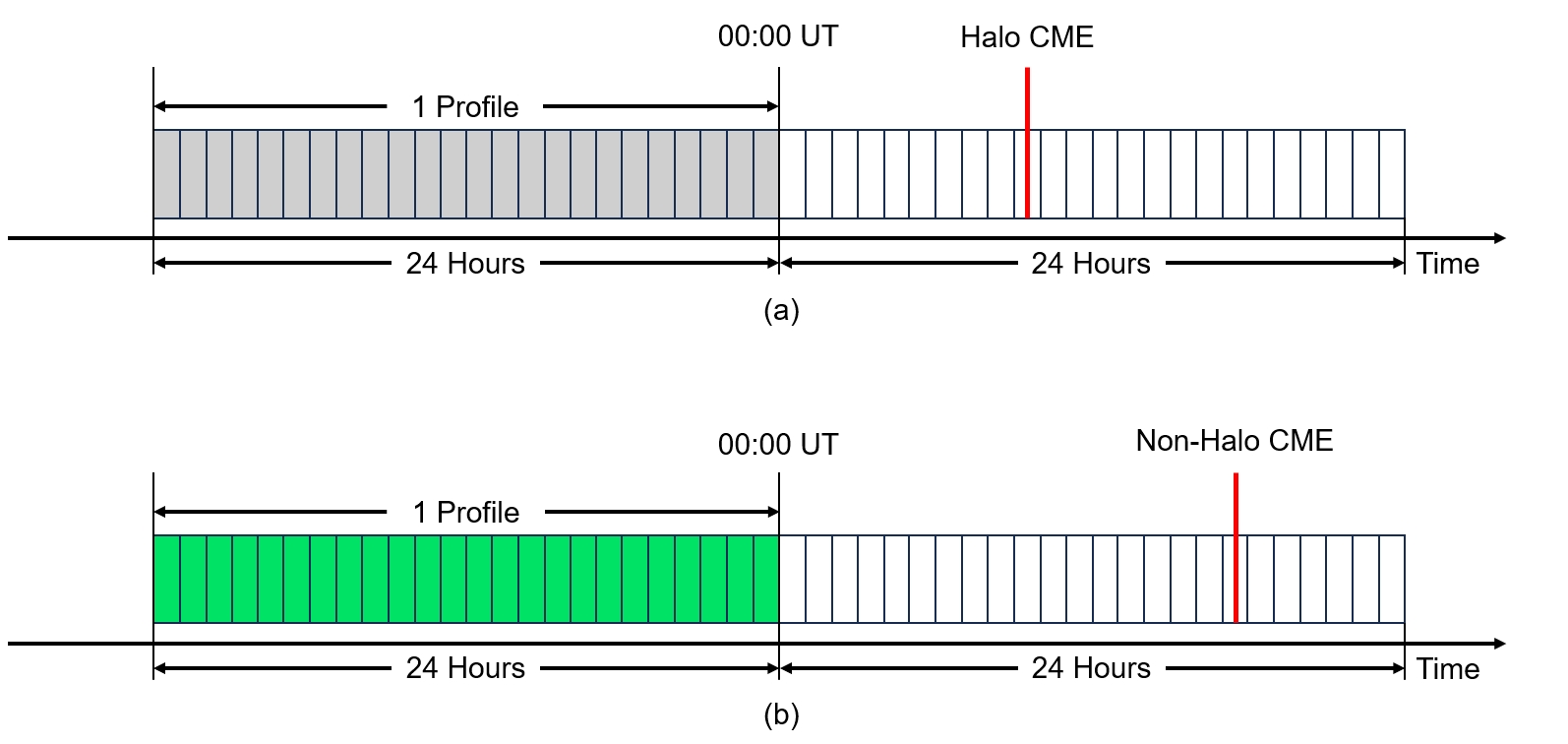}
\caption{Constructing positive and negative data samples in an AR. Data samples are collected at a cadence of 12 minutes. Each rectangular box corresponds to 1 hr and contains five data samples.
(a) Gray data samples collected 24 hours before 
00:00 UT, i.e., 
the beginning of a day,
in which there is a halo CME, are labeled positive. 
These positive data samples are collectively referred to 
as a positive profile.
(b) Green data samples collected 24 hours before 
00:00 UT, i.e., 
the beginning of a day, 
in which there is a 
non-halo CME, are labeled negative.
These negative data samples are collectively referred to as a negative profile.}
\label{fig:label}
\end{figure}

We aim to tackle the following binary classification task.
Given an AR and the time point $t$
at the beginning of a day (00:00 UT),
we predict whether
the AR will produce a halo CME within the next 24 hours of $t$.
A halo CME refers to a full-halo CME or partial-halo CME.
As in previous studies
\citep{2015ApJ...798..135B,
BI2016ApJ,
2020ApJ...890...12L},
data samples
from ARs located outside $\pm$ $70^\circ$ of the center meridian or with incomplete parameters are excluded. 
Figure \ref{fig:label}
explains how we create positive data samples and negative data samples in an AR.
In Figure \ref{fig:label}(a), 
gray data samples collected 24 hours before
the beginning of a day in which there is
a halo CME are labeled positive.
We collectively refer to these gray data samples as a positive profile.
When there are multiple halo CMEs on the same day in an AR,
we collect a positive profile for the first halo CME on that day.
In Figure \ref{fig:label}(b),
green data samples collected 24 hours before
the beginning of a day in which there is
a non-halo CME are labeled negative.
We collectively refer to these green data samples as a negative profile.
When there are multiple non-halo CMEs on the same day in an AR,
we collect a negative profile for the first non-halo CME on that day.
If an AR has a halo CME and a non-halo CME on the same day,
we give priority to the halo CME, 
ignoring the non-halo
CME on that day, and
label the corresponding profile as positive.
For an AR without CMEs,
we collect a profile containing 120 consecutive data samples in the AR and label the profile and data samples as negative.

Thus, the positive class contains positive profiles
collected for halo CMEs while the negative class
contains negative profiles collected for non-halo CMEs and no CMEs.
In our study, there are 158 ARs with halo CMEs, 
153 ARs with non-halo CMEs, 
and 1509 ARs without CMEs.
Ideally, 
with HMI's 12-minute cadence,
there should be 120 data samples in a 24-hour profile.
When more than 50 of the 120 data samples are missing in a profile, the profile is excluded from the study. 
This process yields 249 positive profiles and 1696 negative profiles.
When there are less than 50 missing data samples in a profile,
we use
linear interpolation to
add synthetic data samples with interpolated values
to create a complete non-gapped time-series data set.
Synthetic data
samples are added after normalization of the values of the SHARP parameters,
and hence the synthetic data samples do not affect the normalization procedure. 

We adopt an 80:20 scheme to train and test our DeepHalo model. 
Specifically, we used 80\% of the profiles of each of the positive and
negative classes
for model training and used the remaining 20\% of
the profiles from each class for model testing.
Profiles
in the same AR are placed in the training set or the test set, but not both.
This scheme ensures that our model is trained with data different
from the test data
and makes predictions on the test data that it has never seen
before. 

\newpage
\subsection{Data Preprocessing Algorithms}
\label{subsec:dataProcessing}
Our proposed DeepHalo is a transformer model that
utilizes the attention mechanism to process time series data.
The input to DeepHalo is a profile, which is a time series with 120 data samples.
When there are missing data samples or ``gaps'' in the profile, 
we create synthetic data samples with
interpolated values for all parameters
based on the
linear interpolation method
mentioned in Section \ref{subsec:task}
to fill the ``gaps.''
Our profile data set is imbalanced with
249 positive profiles and 1696 negative profiles,
as described in Section \ref{subsec:task}.
With the 80:20 scheme for model training and testing, there are
198 (51, respectively) positive profiles in the training (test, respectively) set, and 1356 (340, respectively) negative profiles
in the training (test, respectively) set. 
We combat the imbalance issue in the training set
using an oversampling algorithm to
increase the number of positive training profiles
to 1386.
Our oversampling algorithm, described below, is a variant of
the Synthetic Minority Oversampling Technique
\citep[SMOTE;][]{DBLP:journals/jair/ChawlaBHK02}.

Consider each positive profile $P$ in the training set. 
Each data sample in $P$ has 18 SHARP parameters, and therefore each data sample can be treated as a point in the 18-dimensional Euclidean space.
We generate a synthetic profile $P'$ corresponding to $P$ as follows.
For each point/data sample $x = (x_{1}, x_{2}, \ldots , x_{18})$ in $P$, 
we find ten nearest neighbors of $x$ based on the Euclidean distance.
We randomly select one of the ten nearest neighbors, 
denoted by $y = (y_{1}, y_{2}, \ldots , y_{18})$.
We then create a synthetic data sample $z$ corresponding to $x$
where the coordinates of the synthetic data sample
$z = (z_{1}, z_{2}, \ldots , z_{18})$
are calculated by formula:
$z_{i}$ = $x_{i}$ + $\alpha$($y_{i}$ - $x_{i}$),
where $1 \leq i \leq 18$ and
$0 \leq \alpha < 1$ is randomly generated using the
np.random.rand() function from the NumPy library in Python. We label the synthetic data sample $z$ as positive.

Each data sample $x$ in $P$ corresponds to a synthetic data sample $z$, and
these synthetic data samples together constitute the synthetic profile $P'$, which is labeled as positive.
Similarly, we can generate five additional synthetic profiles
that correspond to $P$, all of which are labeled as positive.
In total, we create six synthetic 
positive
profiles for $P$.
There are 198 positive profiles in the training set.
With the newly generated synthetic profiles, we obtain
a total of
198 + (198 $\times$ 6) = 1386
positive profiles in the training set.

\subsection{The DeepHalo Model}
\label{subsec:model}

\begin{figure}
\centering
\includegraphics[width=1\linewidth]{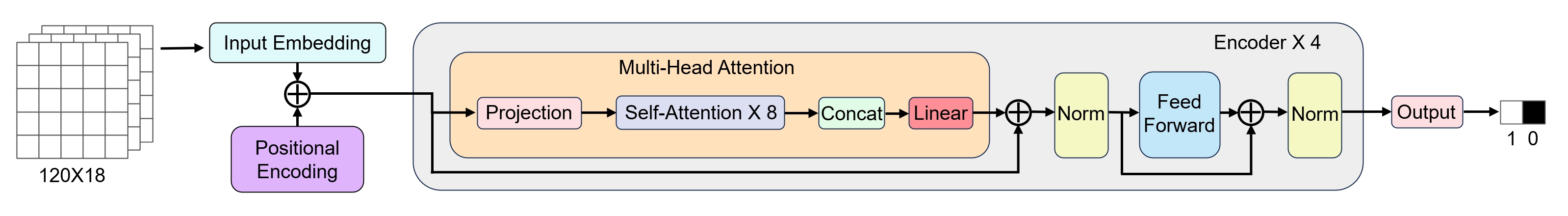}
\caption{Architecture of the proposed DeepHalo model.
The model contains four encoders, where each encoder is composed of a multi-head attention module and a feed-forward neural network. The multi-head attention module has eight heads, 
where each head is a self-attention block. The model accepts as input a profile with 120 data samples where each data sample contains 18 SHARP parameters and produces as output 1 or 0 where 1 indicates that there will be a halo CME within 24 hours and 0 indicates that there will be no halo CME within 24 hours.}
\label{fig:architecture}
\end{figure}

Transformers, originally designed for
natural language processing, 
contain encoders and decoders \citep{DBLP:conf/nips/VaswaniSPUJGKP17}.
DeepHalo only has encoders without decoders
because we process time series, 
not natural languages, 
so there is no need to decode words for sentence translation.
Figure \ref{fig:architecture} shows the architecture of the DeepHalo model.
The model takes as input a profile/time series with 120 data samples
each having 18 SHARP parameters, or features,
of an AR that are collected 24 hours before
the time point $t$ at the beginning of a day
and predicts as output a value of 1 or 0 where
1 indicates that the AR will produce a halo CME within the next
24 hours of $t$ and 0 indicates that
the AR will not produce a halo CME within the next
24 hours of $t$.

In Figure \ref{fig:architecture}, 
the input embedding layer is implemented using a
one-dimensional convolutional
neural network. The convolutional operation transforms the input features into a
higher-dimensional space, 
allowing the DeepHalo model to capture local patterns and
relationships within the input time series effectively.
The positional encoding layer is implemented using a series of sine and cosine functions of
varying frequencies. 
This encoding layer adds information
about the position of each
element in the input sequence, enabling the model to understand and use the
temporal order of the data. 
Thus, the input data undergo the input embedding process first, and then the
positional encoding is added to the embedded representation
of the input data. 
Combining the input embedding and positional encoding
provides a comprehensive representation of each time step
in the input sequence,
encompassing both its feature information and its position information in the sequence.

The encoded input is sent to four encoders, where each encoder is composed of a multi-head attention module and a feed forward neural network with 256 neurons.
The multi-head attention module
consists of a projection layer, followed by eight heads where each head is a self-attention block, followed by a concatenation layer that concatenates the output data from the eight heads.
A linear transformation layer is then applied to these concatenated data to ensure structural consistency between the output and the input.
The self-attention mechanism helps the model determine which data samples in the input time series should receive more attention.
Each head computes attention independently, enabling the model to simultaneously focus on different facets of the input data. This independent computation is pivotal in allowing the model to discern and integrate a broad range of features and dependencies.
The multi-head attention mechanism enhances the model's ability to process the input time series by capturing more complex dependencies of its elements/data samples in a more nuanced manner.

Overall, the multi-head attention module is devoted to calculating the input's self-attention in different heads. 
The feed-forward neural network allows for additional refinement of the data processed by the self-attention mechanism. 
The multi-head attention module (feed forward neural network, respectively) includes a residual connection, followed by a normalization layer. 
Residual connections ensure information preservation, while normalization layers stabilize feature distribution, aiding in faster model convergence.
Finally, the output layer produces a predicted value of 1 or 0 where 1 indicates that there will be a halo CME within 24 hours and 0 indicates that there will be no halo CME within 24 hours.

During training, a weighted cross-entropy loss function is utilized to
optimize the model parameters. 
This loss function $\cal L$ is defined as:
\begin{equation}
{\cal L} = \sum_{n=1}^{N} \omega_0 (1 - y_n) \log (1 - \hat{y}_n)  +  \omega_1 y_n \log (\hat{y}_n).
\end{equation}
Here, $N$ = 2742
denotes the total number of
profiles with 1386 positive and 1356 negative profiles, 
each comprising
120 consecutive data samples, in the training set. 
The weights \( \omega_0 \) and \( \omega_1 \) correspond to the negative and positive classes, respectively. These weights are derived on the basis of the class size ratio.
The observed probability \( y_n \) is set to 1 if the 
\( n \)th 
profile
belongs to the positive class, or set to 0 otherwise. 
The estimated probability 
\( \hat{y}_n \) of the \( n \)th 
profile
is computed by the model. 
The optimizer used is 
Adam \citep{DBLP:journals/corr/KingmaB14}, 
which is a stochastic gradient descent method, with a learning rate of 0.001. 

\section{Results} 
\label{sec:results}

\subsection{Evaluation Metrics} 
\label{subsec:metrics}

Given an AR and a test profile $P$ 
of the time point $t$ at the beginning of a day
where $P$ contains the data samples collected 24 hours before $t$ in the AR,
we define $P$ as a true positive (TP) 
if our DeepHalo model predicts that $P$ is positive,
that is, the AR will produce a halo CME within the next 24 hours of $t$,
and the profile $P$ is indeed positive.
We define $P$ as a false positive (FP)
if our model predicts that $P$ is positive
while $P$ is actually negative, i.e., 
the AR will not produce a halo CME within the next 24 hours of $t$.
We say $P$ is a true negative (TN) if our model
predicts that $P$ is negative
and $P$ is indeed negative; 
$P$ is a false negative (FN) if
our model predicts that $P$ is
negative while $P$ is actually positive.
When the context is clear, we also use
TP (FP, TN, and FN, respectively) 
to represent the number of
true positives
(false positives, true negatives, and false negatives, respectively) 
produced by our model.

The evaluation metrics used in this study include the
following:
\begin{equation}
    \text{Recall} = \frac{\mbox{TP}}{\mbox{TP} + \mbox{FN}},
\end{equation}

\begin{equation}
    \text{Precision} = \frac{\mbox{TP}}{\mbox{TP} + \mbox{FP}},
\end{equation}

\begin{equation}
    \text{ACC} = \frac{\mbox{TP} + \mbox{TN}}
    {\mbox{TP} + \mbox{FP} + \mbox{TN} + \mbox{FN}},
\end{equation}

\begin{equation}
\text{F1} =  \frac{2 \times \text{TP}}{2 \times \text{TP} + \text{FP}+\text{FN}},
\end{equation}

\begin{equation}
    \text{HSS} = \frac{2 \times 
    (\mbox{TP} \times \mbox{TN} - 
    \mbox{FP} \times \mbox{FN})}
    {(\mbox{TP} + \mbox{FN}) \times 
    (\mbox{FN} + \mbox{TN}) + 
    (\mbox{TP} + \mbox{FP}) \times 
    (\mbox{FP} + \mbox{TN})},
\end{equation}

\begin{equation}
    \text{TSS} = \frac{\mbox{TP}}
    {\mbox{TP} + \mbox{FN}} - 
    \frac{\mbox{FP}}{\mbox{FP} + \mbox{TN}}.
\end{equation}

We compute TP, FP, TN and FN in the test set.
Our test set of 391 profiles
is imbalanced in that it contains
51 positive test profiles and 340 negative test profiles,
with the number of negative test profiles
far exceeding the number of positive test profiles.
ACC is not suitable for imbalanced classification because a naive classifier predicting all
test profiles in the minority/positive class to 
belong to the majority/negative class
would still get a high ACC value.
On the other hand,
the F1 score is a commonly used evaluation metric in binary classification tasks, providing a balanced measure of the performance of a classifier
\citep{2024ApJS..271...29A}.
The Heidke skill score
\citep[HSS;][]{1b22346b-53d4-3dc7-a5a2-a35717d56059} measures the fractional
improvement in the prediction of a model over the random prediction.
The true skill statistics (TSS) score
\citep{2012ApJ...747L..41B},
widely used in imbalanced solar eruption predictions
\citep{2020ApJ...890...12L,2023NatSR..1313665A,2024ApJS..271...29A},
is the difference between the recall rate and
the false alarm rate.
We will mainly focus on TSS
where the larger the TSS score a
method has, the better performance the method achieves.

\begin{figure}
\centering
\includegraphics[width=1\linewidth]{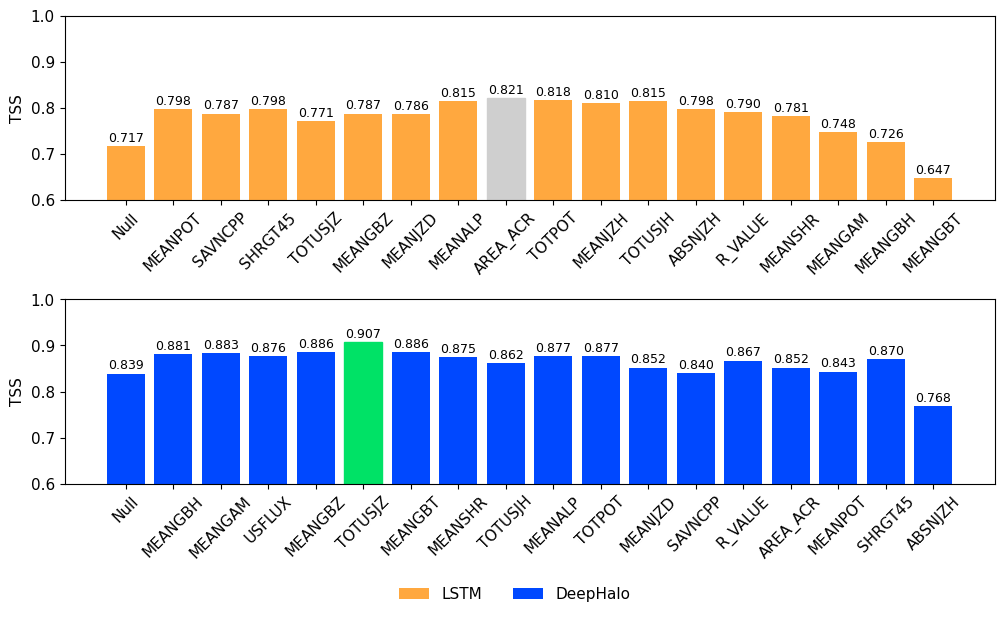}
\caption{Analysis of 
feature contributions and selection of best features for the LSTM (top) and DeepHalo (bottom) models.
There are 18 SHARP parameters in total.
The best LSTM performance
is achieved by removing 8 parameters 
(MEANPOT, SAVNCPP, SHRGT45, TOTUSJZ, MEANGBZ, MEANJZD, MEANALP, and AREA\_ACR)
and only using the remaining 10 parameters,
with a TSS of 0.821, as highlighted by the gray bar 
in the LSTM chart at the top.
The best DeepHalo performance
is achieved by removing 5 parameters 
(MEANGBH, MEANGAM, USFLUX, MEANGBZ, and TOTUSJZ)
and only using the remaining 13 parameters,
with a TSS score of 0.907, as highlighted
by the green bar in the DeepHalo chart at the bottom.
}
\label{fig:feature_contribution_rank}
\end{figure}

\subsection{Feature Analysis}
We first analyze the contributions of the 18 SHARP
parameters or features described in Section
\ref{sec:data} to the DeepHalo model
using a recursive parameter elimination algorithm
\citep{FeatureEngineeringSelection} 
to select a set of parameters that achieve the best
performance, where the performance is measured by TSS. 
The algorithm selects parameters by recursively considering smaller and
smaller sets of parameters, where the most negatively contributing or
the least positively contributing parameters
are successively pruned from the current sets of parameters.
Initially, in iteration 0, the model is trained using the set of all 18 SHARP parameters and the TSS score is calculated, which is the largest TSS recorded in iteration 0.
Then, in iteration $i$, $1 \leq i \leq 17$, 
each parameter in the current set is removed in turn with replacement.
After a parameter is removed, the model is re-trained using the remaining parameters, and the re-trained model is used to make predictions on the test set to calculate its TSS score.
The largest TSS score thus obtained in iteration $i$ and the corresponding parameter (that is, the removed parameter $f$) are recorded.
There are two cases.
In case 1, the largest TSS recorded in iteration $i$ is
greater than or equal to the largest TSS recorded in iteration $i-1$.
This means that the removed parameter $f$ in iteration $i$
has the most negative contribution to the model in iteration $i-1$. 
In case 2, the largest TSS recorded in iteration $i$ is less than
the largest TSS recorded in iteration $i-1$. 
This means that the removed parameter $f$ in iteration $i$
has the least positive contribution to the model in iteration $i-1$.
We remove the parameter $f$ from the parameter set in iteration $i$ and enter iteration $i+1$
in which the parameter set is updated to contain the parameters
in iteration $i$ minus the
removed parameter $f$.

Figure \ref{fig:feature_contribution_rank} presents the results of the parameter/feature contribution analysis.
Since LSTM networks have been used in CME
prediction, we also included an LSTM network in our study. 
The details of the configuration of the LSTM network
can be found in
\citet{2019ApJ...877..121L,2020ApJ...890...12L}.
As in DeepHalo, the LSTM network adopts the weighted cross-entropy loss
function and the Adam optimizer to minimize the loss. The SMOTE-based algorithm was also used by the LSTM model to increase the
number of positive training profiles so that the size of the positive training set is approximately the same as the size
of the negative training set.

Referring to the 
DeepHalo model in
Figure \ref{fig:feature_contribution_rank},
we see that MEANGBH makes a negative contribution because removing it increases the TSS from 0.839 to 0.881, while
ABSNJZH makes a positive contribution, since removing it decreases the TSS from 0.870 to 0.768.
Furthermore, comparing the charts for LSTM and DeepHalo
in Figure \ref{fig:feature_contribution_rank},
we see that
the parameters are eliminated in different ways.
For example, for the LSTM model,
the remaining parameter in iteration 18
(not shown in the LSTM chart at the top) is USFLUX.
However, for the DeepHalo model,
the remaining parameter in iteration 18 (not shown in the DeepHalo chart at the bottom)
is MEANJZH.
The best LSTM performance
is achieved by removing 8 parameters 
(MEANPOT, SAVNCPP, SHRGT45, TOTUSJZ, MEANGBZ, MEANJZD, MEANALP, and AREA\_ACR)
and only using the remaining 10 parameters,
with a TSS of 0.821, as highlighted by the gray bar in the LSTM chart.
The best DeepHalo performance
is achieved by removing 5 parameters 
(MEANGBH, MEANGAM, USFLUX, MEANGBZ, and TOTUSJZ)
and only using the remaining 13 parameters,
with a TSS of 0.907, as highlighted
by the green bar in the DeepHalo chart.
As a result, we
adopted the best-performing LSTM with 10 SHARP parameters 
and the best-performing DeepHalo with 13 SHARP parameters in subsequent experiments.
It is worth pointing out that 
TOTUSJH (the complete unsigned current helicity), 
TOTPOT (the total photospheric magnetic free energy density),
and R\_VALUE (the accumulated flux near the polarity inversion line)
are among the 10 parameters used by LSTM and
the 13 parameters used by DeepHalo,
indicating that they are important SHARP parameters
for solar eruption prediction.
This finding is consistent with those
reported in the literature
\citep{2015ApJ...798..135B,2017ApJ...843..104L}.

\begin{figure}
\centering
\includegraphics[width=0.9\linewidth]{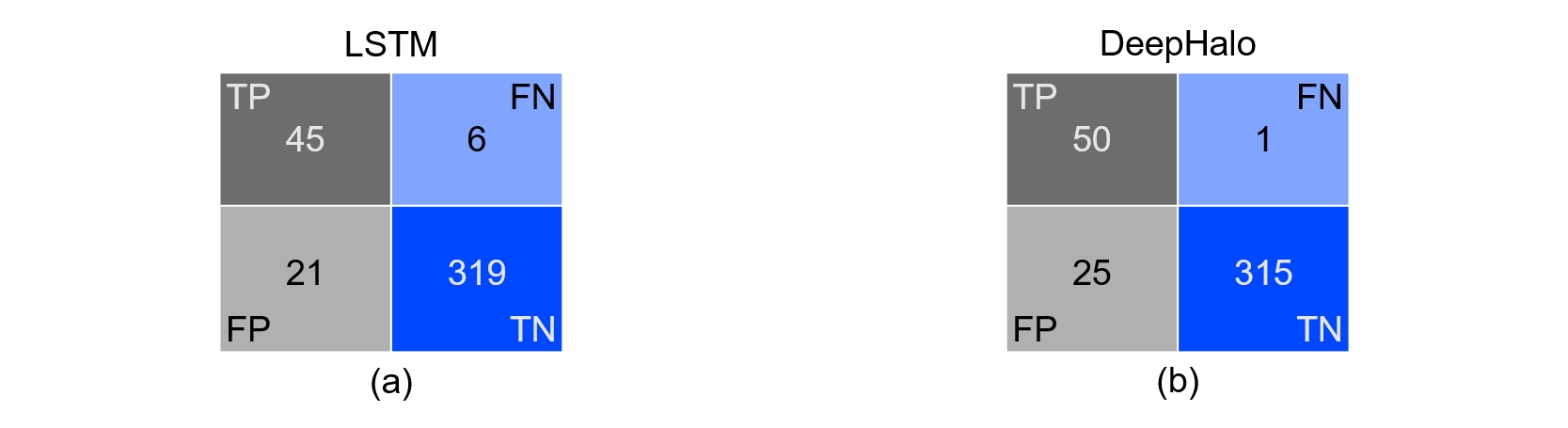}
\caption{Confusion matrices of (a) LSTM and (b) DeepHalo based on the test set used in our study.
}
\label{fig:confusion1fold}
\end{figure}

\begin{figure}
\centering
\includegraphics[width=1\linewidth]{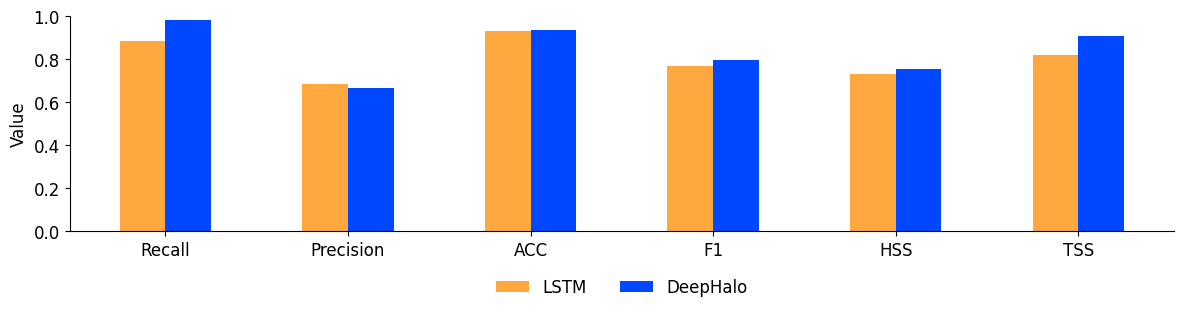}
\caption{Bar graphs showing the comparison between LSTM and DeepHalo based on the test set used in our study.
}
\label{fig:comparisonresults}
\end{figure}

\subsection{Performance Assessment}
\label{subsec:assessment}
Figure \ref{fig:confusion1fold} presents the confusion matrices 
for the best-performing LSTM and the best-performing DeepHalo models.
It can be seen in Figure \ref{fig:confusion1fold}
that DeepHalo is more sensitive than LSTM
in the sense that
DeepHalo has a higher FP than LSTM.
However, LSTM misses more halo CMEs than DeepHalo, having a
higher FN than DeepHalo.
Figure \ref{fig:comparisonresults}
compares the performance metric values of the two deep learning models.
The results in Figure \ref{fig:comparisonresults}
are consistent with
the confusion matrices in
Figure \ref{fig:confusion1fold}.
We see that LSTM has a higher precision than DeepHalo,
while DeepHalo has a higher recall than LSTM. 
DeepHalo achieves a TSS score of
0.907, which is
better than the TSS score of
0.821 obtained by LSTM.

We adopted the SMOTE-based algorithm
to combat the imbalance issue in the training set.
One wonders how effective this algorithm is.
We conducted an additional experiment
in which we ran DeepHalo on the original imbalanced training set without using the SMOTE-based algorithm.
We denote this
version in which the SMOTE-based algorithm is not used as
DeepHalo-Imbalance,
and compare it with DeepHalo.
As described in Section \ref{subsec:dataProcessing},
the original training set has 198 positive training profiles and 1356 negative training profiles, totaling 1554 training profiles.
The test set remains the same, 
with 51 positive test profiles
and 340 negative test profiles, totaling 391 test profiles.
Our experimental results show that
DeepHalo with a TSS score of
0.907
is better than DeepHalo-Imbalance 
with a TSS score of
0.873.
The SMOTE-based algorithm improves DeepHalo's performance, 
although DeepHalo-Imbalance 
also works reasonably well
due to the weighted cross-entropy loss function used by the model.
This loss function takes into account
the class size ratio
between the positive/minority class and the negative/majority class, 
helping to combat the imbalance issue in the training set.

\subsection{Model Interpretation}

\begin{figure}
\centering
\includegraphics[width=0.9\linewidth]{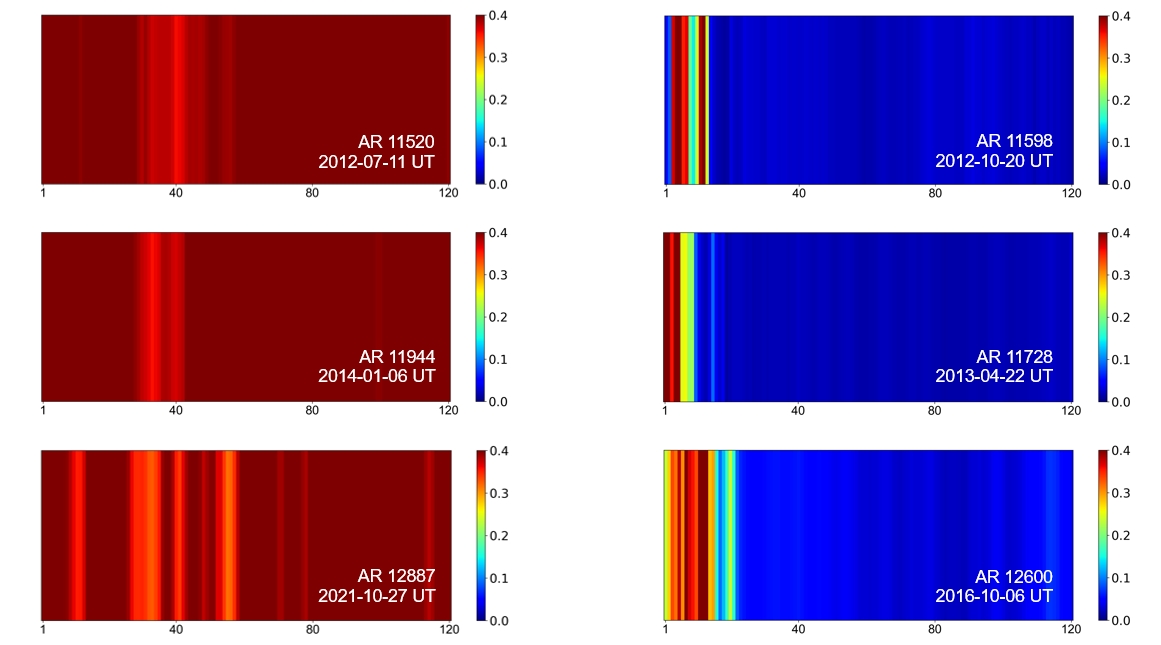}
\caption{Attention heatmaps of the DeepHalo model
for six test profiles along with their collection dates in six ARs, respectively.
In each heatmap, the X-axis represents data samples in the corresponding test profile and
the Y-axis denotes the scaled attention score,
represented by a color, for each data sample in the profile. 
A larger attention score at a data sample on the X-axis 
indicates more attention is 
paid to the data sample, where 
large attention scores are represented by dark red
and small attention scores are represented by dark blue.
Attention scores on the six color bars 
for the six test profiles are scaled differently to 
facilitate visualization and to better show the attention distribution on the 120 data samples in each test profile.
(Left panels) 
Attention heatmaps for three positive predictions
where a positive prediction indicates that 
the corresponding AR will produce a halo CME within the next 24 hours.
(Right panels) 
Attention heatmaps for three negative predictions
where a negative prediction indicates that the corresponding AR will not produce a halo CME within the next 24 hours.
The attention heatmaps on the left display a relatively uniformly distributed pattern, while the heatmaps on the right show more focused attention on the first several data samples in a test profile.}
\label{fig:halo_heatmap}
\end{figure}

To understand and explain a prediction made by our DeepHalo model,
we computed the average attention heatmap
of the final layer of the multi-head attention module
in DeepHalo's final (i.e., the fourth) encoder
\citep{DBLP:journals/tvcg/YehCWCVW24}.
This pivotal layer, epitomizing the culmination of preceding transformations, forms intricate high-level representations critical for the model's decision-making process. 
A thorough exploration of this layer's attention distribution uncovers essential insights into the model's priorities for predicting a halo CME, thereby enriching our comprehension of
the model's focus in making an inference.
Specifically, the average attention heatmap
sheds light on key data samples in an input profile that are crucial to the prediction made by the model. 
The interpretive analysis of the heatmap offers a direct visual representation of the model's attention allocation across the data samples in the input profile.

Figure \ref{fig:halo_heatmap} presents 
heatmaps for 
six test profiles along with their collection dates in six ARs, respectively.
The left panels display the heatmaps for three positive predictions,
where a positive prediction indicates that the corresponding AR will produce a halo CME within the next 24 hours.
These heatmaps show a relatively uniformly distributed attention pattern. 
This uniformity indicates DeepHalo's perception
of the halo CME emergence as a progressive process, consistent with the dynamic and evolution nature of the AR that emits
the halo CME.
The right panels display the heatmaps for three negative predictions,
where a negative prediction indicates that the corresponding AR will not produce a halo CME within the next 24 hours.
These heatmaps show more focused attention, particularly around crucial time
intervals near the prediction point (in these instances, the model’s attention is mainly paid to the first several data samples
in a test profile). This focused attention suggests the ability of the model to identify temporal markers indicative of a
negative prediction, underscoring the significance of specific time segments in the prediction. Moreover, the heatmap
highlights intervals that are less critical, getting less attention, in the prediction.

In summary, the attention heatmaps
can be interpreted as a consequence of the model's analysis of the input profile. 
If the beginning of the input profile shows no promise for a halo CME to occur, then the model more or less discards the rest of the input profile. 
If it shows promise for a halo CME to occur, the model considers
the data samples till the end to see if the input profile indicates whatever process leads to the ejection of the halo CME. 
While there is lacking a 
theory that relates the
model's attention with the occurrence of a halo CME, 
these heatmaps might be able to help investigate the physical nature of the process leading up to such an event. 

\section{Discussion and Conclusion} 
\label{sec:conclusion}

\begin{figure}
\centering
\includegraphics[width=0.9\linewidth]{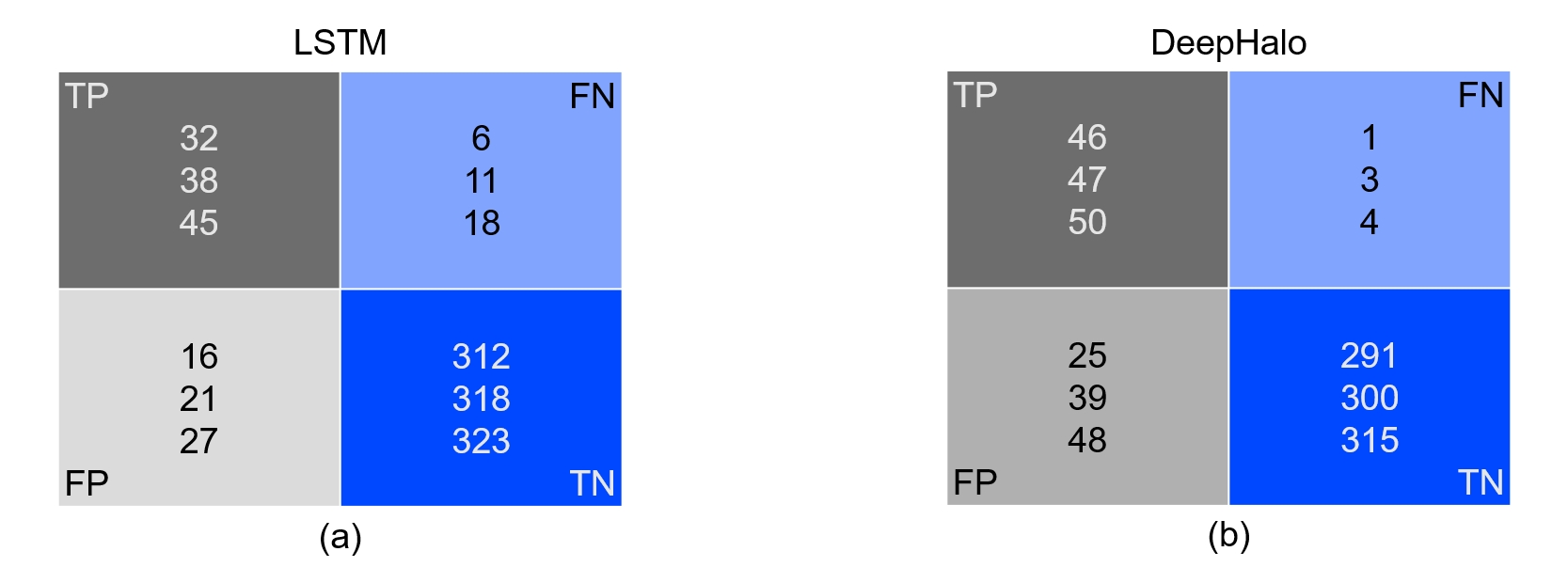}
\caption{Confusion matrices of (a) LSTM and (b) DeepHalo obtained from 
five-fold cross validation. The figure displays (from top to bottom) the minimum, medium and maximum TP, FN, TN, and FP, respectively, from the five runs based on the five-fold cross validation scheme.}
\label{fig:confusion5fold}
\end{figure}

We developed a transformer model (DeepHalo) to predict whether an AR would produce a halo CME within the next 24 hours of $t$ where $t$ denotes the beginning of a day.
The model takes as input the profile $P$ of $t$ where $P$
contains the data samples collected 24 hours before $t$ in the AR
and outputs ``yes'' indicating that there will be a halo CME within the
next 24 hours of $t$ or ``no'' indicating that there will be no halo CME within the next 24 hours of $t$.
Each data sample initially contains 18 SHARP parameters downloaded from the JSOC website.
We adopted an 80:20 scheme to train and
test our DeepHalo model and compared the model
with a closely related LSTM network previously used for CME prediction
\citep{2020ApJ...890...12L}.
Experimental results showed that DeepHalo
using only 13 SHARP parameters performs the best
with a TSS score of 0.907,
while the LSTM network using only 10 SHARP parameters performs the best with a TSS score of 0.821.
Hence, for each model, we use its optimal SHARP parameters in our study.
The fact that DeepHalo outperforms the LSTM network in TSS
is mainly attributed to the high recall rate of
DeepHalo, which detects 50 of 51 halo CMEs
in the test set.
However, the LSTM network only detects
45 of the 51 test halo CMEs.

To explain a prediction made by the DeepHalo model, we implement the average attention heatmap
of the final layer of the multi-head attention module
in DeepHalo's final (i.e., the fourth) encoder.
This heatmap sheds light on key data samples in a test profile that are crucial to the prediction.
The heatmap displays a relatively uniformly distributed attention pattern for a positive prediction (that is, a prediction indicating that there will be a halo CME within the next 24 hours of $t$).
However, for a negative prediction
(that is, a prediction indicating that there will be no halo CME within the next 24 hours of $t$),
the heatmap shows more focused attention.
The interpretive analysis of the heatmap offers a direct visual representation of the model's attention allocation across the data samples
in the test profile and
helps one better understand the model's decision-making process.

The training set used in our study is imbalanced in the sense that
negative training profiles
outnumber positive training profiles.
To combat the imbalance issue in the training set,
we employ a SMOTE-based algorithm to
create synthetic positive training profiles
so that the positive training set has roughly
the same size as the negative training set.
The experimental results
showed that the DeepHalo model combined with the SMOTE-based algorithm improves the performance of the model.
Specifically, DeepHalo combined with the
SMOTE-based algorithm achieves a TSS score of
0.907 
compared to the TSS score of
0.873 
obtained by the model without using
the SMOTE-based algorithm.

\begin{figure}
\centering
\includegraphics[width=1\linewidth]{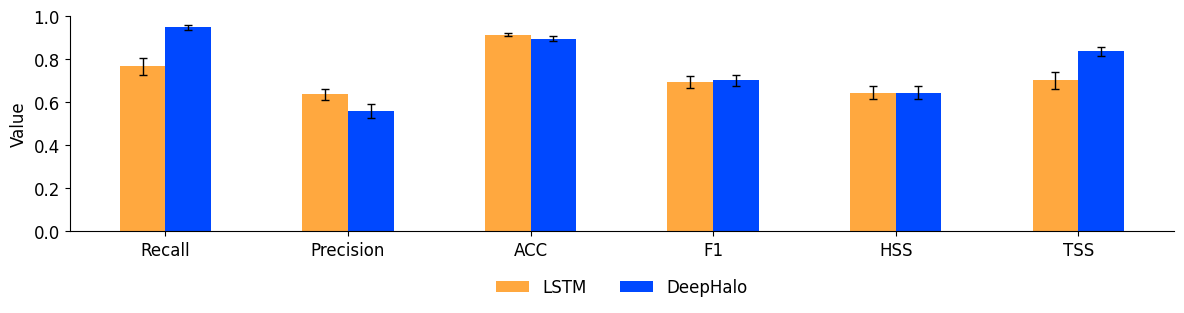}
\caption{Bar graphs showing the comparison between LSTM and DeepHalo
based on the five-fold cross validation scheme.
}
\label{fig:comparisonresults5fold}
\end{figure}

To further understand the behavior of our DeepHalo model and
compare it with the related LSTM network,
we conducted an additional five-fold cross-validation experiment.
Specifically, we divided the profile data set used in the study
into five equally sized distinct partitions or folds.
Profiles from the same AR were placed in the same fold.
Each two folds had roughly the same number of
positive profiles (negative profiles, respectively).
In the run $i$, where $1 \leq i \leq 5$, 
we used the fold $i$ as the test set and
the union of the other four folds as the training set.
The prediction accuracy in each run was calculated and the mean
and standard deviation over the five runs were plotted.
Figure \ref{fig:confusion5fold} presents the confusion matrices
of the LSTM and DeepHalo.
Figure \ref{fig:comparisonresults5fold}
compares the performance metric values of the two deep learning models
where each colored bar represents the mean of the five runs and its associated
error bar represents the standard deviation divided by the
square root of the number of runs.
DeepHalo achieves a mean TSS score of
0.836,
which is better than the mean TSS score of 
0.702 obtained by the LSTM network.
This result is consistent with that of Figure \ref{fig:comparisonresults},
demonstrating the superiority of DeepHalo over
the closely related LSTM method.
On the basis of the results, we conclude that DeepHalo
is a feasible tool for predicting the occurrence of halo CMEs.

\begin{acknowledgments}
The authors thank members of the Institute for Space Weather Sciences
for fruitful discussions.
We also thank the team of SDO/HMI for providing vector magnetic field data products. 
The CME event records were retrieved from the DONKI
and LASCO CME Catalog, which was created and maintained at the CDAW Data Center by NASA and the Catholic University of America
in cooperation with the Naval Research Laboratory.  
The deep learning methods studied here
were implemented in PyTorch.
J.J. acknowledges support from NSF grants
AGS-2149748 and AGS-2300341.
J.W. and H.W. acknowledge support from NSF grants
AGS-2149748, AGS-2228996, OAC-2320147, and NASA grants 80NSSC24K0548,
80NSSC24K0843, and 80NSSC24M0174.
Y.X. acknowledges support from NSF grants
AGS-2228996,
AGS-2229064, and RISE-2425602.
K.A. acknowledges support from King Saud University, Saudi Arabia.
V.Y. acknowledges support from NSF grants
AST-2108235,
AGS-2114201,
AGS-2300341, and
AGS-2309939.
\end{acknowledgments}

\bibliographystyle{aasjournal}

\end{document}